\documentclass[iop]{emulateapj}
\usepackage{graphicx}
\usepackage{amssymb}
\usepackage{epstopdf}
\usepackage{amsmath}
\usepackage[breaklinks]{hyperref}
\usepackage{ulem}

\newcommand{\ms}{\ensuremath{\rm m\,s^{-1}}}

\newcommand{\gcmc}{\ensuremath{\rm g\,cm^{-3}}}
\newcommand{\gcc}{\gcmc}

\newcommand{\teff}{\ensuremath{T_{\rm eff}}}

\newcommand{\rsun}{\ensuremath{R_\sun}}

\newcommand{\rstar}{\ensuremath{R_\star}}

\newcommand{\rpl}{\ensuremath{R_{\rm P}}}
\newcommand{\mpl}{\ensuremath{M_{\rm P}}}

\newcommand{\rhopl}{\ensuremath{\rho_{\rm P}}}

\newcommand{\rearth}{\ensuremath{R_\earth}}
\newcommand{\mearth}{\ensuremath{M_\earth}}
\newcommand{\fearth}{\ensuremath{F_\earth}}

\newcommand{\rspecial}{4 \rearth}

\begin{document}
\title{The Mass-Radius Relation for 65 Exoplanets Smaller than 4 Earth Radii}
\author{Lauren~M.~Weiss$^{1,\dagger}$ \& Geoffrey~W.~Marcy$^1$}
\affil{$^1$B-20 Hearst Field Annex, Astronomy Department, University of California, Berkeley, CA 94720}
\altaffiltext{$\dagger$}{\small Supported by the NSF Graduate Student Fellowship, Grant DGE 1106400.}

\submitted{Accepted to ApJ Letters on 30 January, 2013.} 

\begin{abstract}
We study the masses and radii of 65 exoplanets smaller than 4\rearth\ with orbital periods shorter than 100 days.  We calculate the weighted mean densities of planets in bins of 0.5 \rearth\ and identify a density maximum of 7.6 \gcc at 1.4 \rearth.  On average, planets with radii up to $\rpl=1.5\rearth$ increase in density with increasing radius.   Above 1.5 \rearth, the average planet density rapidly decreases with increasing radius, indicating that these planets have a large fraction of volatiles by volume overlying a rocky core.  Including the solar system terrestrial planets with the exoplanets below 1.5 \rearth, we find $\rhopl = 2.43 + 3.39 \left(\rpl/\rearth\right) \gcc$ for $\rpl < 1.5 \rearth$, which is consistent with rocky compositions.  For $1.5 \le \rpl/\rearth < 4$, we find $\mpl/\mearth = 2.69 \left(\rpl/\rearth\right)^{0.93}$.  The RMS of planet masses to the fit between 1.5 and 4 \rearth\ is 4.3 \mearth\ with reduced $\chi^2=6.2$.  The large scatter indicates a diversity in planet composition at a given radius.  The compositional diversity can be due to planets of a given volume (as determined by their large H/He envelopes) containing rocky cores of different masses or compositions.
\end{abstract}


\section{Introduction}

The Kepler Mission has found an abundance of planets with radii $\rpl < 4\rearth$ \citep{Batalha2013}; the most recent head-count indicates 3206 planet candidates in this size range (NASA exoplanet Archive, queried 15 Jan. 2014), most of which are real \citep{MortonJohnson2011}.  Although there are no planets between the size of Earth and Neptune in the solar system, occurrence calculations that de-bias the orbital geometry and completeness of the Kepler survey find that planets between the size of Earth and Neptune are common in our galaxy, occurring with orbital periods between 5 and 50 days around 24\% of stars \citep{Petigura2013a}.  However, in many systems, it is difficult to measure the masses of such small planets because the gravitational acceleration these planets induce on their host stars or neighboring planets is challenging to detect with current telescopes and instruments.  We cannot hope to measure the masses of all planets in this size range discovered by Kepler.  Obtaining measurements of the masses of a subset of these planets and characterizing their compositions is vital to understanding the formation and evolution of this population of planets.

Many authors have explored the relation between planet mass and radius as a means for understanding exoplanet compositions and as a predictive tool.  \citet{Seager2007} predict the mass-radius relationship for  planets of various compositions.  The mass-radius relation in \citet{Lissauer2011}, which is commonly used in literature to translate between planet masses and radii, is based on fitting a power law relation to Earth and Saturn only.  Other works, such as \citet{Enoch2012, Kane2012, Weiss2013}, determine empirical relations between mass and radius based on the exoplanet population.  

Recent mass determinations of small planets motivate a new empirical mass-radius relation.  Restricting the empirical mass-radius relation to small exoplanets will improve the goodness of fit, allowing better mass predictions and enabling a superior physical understanding of the processes that drive the mass-radius relation for small planets.  

One challenge in determining a mass-radius relation for small planets is the large scatter in planet mass.  At 2\rearth, planets are observed to span a decade in density, from less dense than water to densities comparable to Earth's.  This scatter could result from measurement uncertainty or from compositional variety among low-mass exoplanets.

In this paper, we investigate mass-radius relationships for planets smaller than 4 Earth radii.  We explore how planet composition--rocky versus rich in volatiles-- influences the mass-radius relationship.  We also investigate the extent to which system properties contribute to the scatter in the mass-radius relation by examining how these properties correlate with the residuals of the mass-radius relation.


\section{Selecting Exoplanets with Measured Mass and Radius}
We present a judicious identification of small transiting planets with masses or mass upper limits measured via stellar radial velocities (RVs) or numerically modeled transit timing variations (TTVs).  The only selection criterion was that the exoplanets have $\rpl < 4 \rearth$ and either a mass determination, a marginal mass determination, or a mass upper limit.  There were no limits on stellar type, orbital period, or other system properties.

We include all 19 planets smaller than 4\rearth\ with masses vetted on exoplanets.org, as of January 13, 2013.  Twelve of these masses are determined by RVs, but the masses of four Kepler-11 planets, Kepler-30 b, and two Kepler-36 planets are determined by TTVs \citep{Lissauer2013, Sanchis-Ojeda2012, Carter2012}.  We include five numerically-determined planet masses from TTVs not yet on exoplanets.org: three KOI-152 (Kepler-79) planets \citep{Jontof-Hutter2013}, and two KOI-314 planets \citep{Kipping2014}.  We also include all 40 transiting planets with RV follow-up in \citet{Marcy2014} that are smaller than 4\rearth, and the RV-determined mass of KOI-94 b \citep{Weiss2013}, none of which yet appear on exoplanets.org.  

\begin{figure*}[htbp] 
   \centering
   \includegraphics[width=7in]{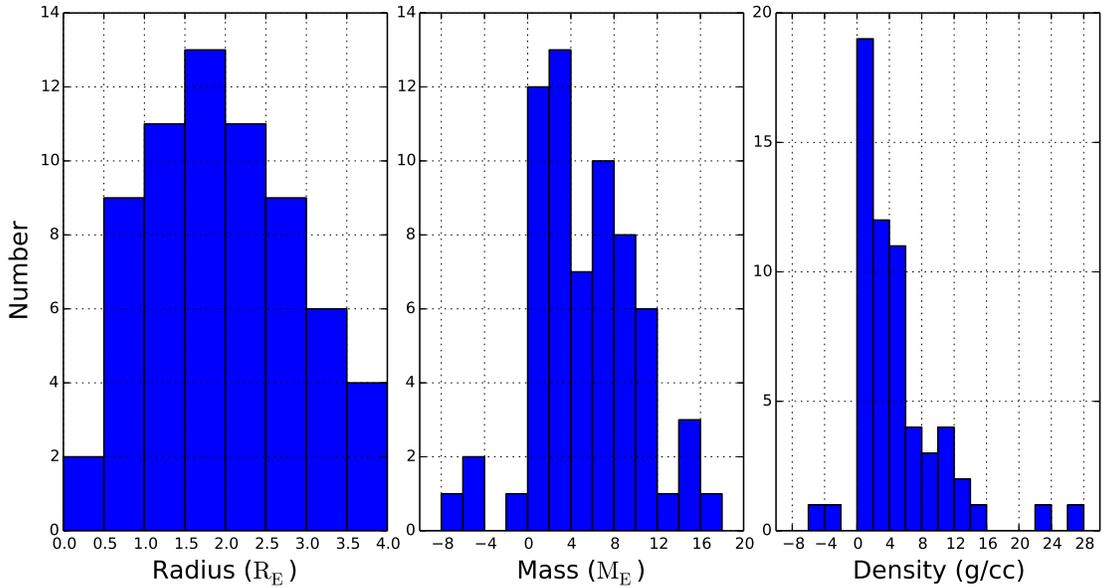} 
   \caption{\small Histograms of exoplanet radii, masses, and densities for the 65 exoplanets smaller then 4 Earth radii with measured masses or mass upper-limits.  Extreme density outliers Kepler-37 b, Kepler-100 d, Kepler-106 c, and Kepler-131 c are excluded from the density histogram for clarity, but are included in Table 1 and the fits.}
\label{fig:histograms}
\end{figure*}

55 Cnc e, Corot-7 b, and GJ 1214 b have been studied extensively, and we had to choose from the masses and radii reported in various studies.  For 55 Cnc e, we use $\mpl = 8.38\pm0.39$, $\rpl = 1.990\pm0.084$ \citep{Endl2012,Dragomir2013a}; for Corot-7 b, we use $\mpl =7.42\pm1.21$, $\rpl= 1.58\pm 0.1$ \citep{Hatzes2011}, and for GJ 1214 b, we use $\mpl =6.45\pm0.91 $, $\rpl= 2.65\pm0.09$ \citep{Carter2011}.  Histograms of the distributions of planet radius, mass, and density are shown in Figure \ref{fig:histograms}, and the individual measurements of planet mass and radius are listed in Table \ref{tab:mrf}.

The exoplanets all have $P < 100$ days.  This is because the transit probability is very low for planets at long orbital periods and because short-period planets are often favored for RV and TTV studies.

\begin{deluxetable*}{lllllll}


\tabletypesize{\tiny}

\tablewidth{0pt} 

\tablecaption{Exoplanets with Masses or Mass Upper Limits and $\rpl < \rspecial$}
\tablenum{1}

\tablehead{\colhead{Name} & \colhead{Per} & \colhead{Mass} & \colhead{Radius} & \colhead{Flux$^a$} & \colhead{First Ref.} & \colhead{Mass, Radius Ref.} \\ 
\colhead{} & \colhead{(d)} & \colhead{($\mearth$)} & \colhead{($\rearth$)} & \colhead{($\fearth$)} & \colhead{} & \colhead{} } 

\startdata
           $^b$55 Cnc e &      0.737 &       8.38$\pm$0.39       &       1.990$\pm$0.084       &   2400 &                     \citet{McArthur2004} &               \citet{Endl2012},\\ 
           & & & & & & \citet{Dragomir2013a} \\
           CoRoT-7 b &      0.854 &       7.42$\pm$1.21       &       1.58$\pm$0.1       &   1800 &             \citet{Queloz2009}, &                       \citet{Hatzes2011}\\ 
            & & & & & \citet{Leger2009} & \\
           GJ 1214 b &      1.580 &       6.45$\pm$0.91       &       2.65$\pm$0.09       &     17 &                  \citet{Charbonneau2009} &                       \citet{Carter2011}\\ 
          HD 97658 b &      9.491 &       7.87$\pm$0.73       &       2.34$\pm$0.16       &     48 &                       \citet{Howard2011} &                     \citet{Dragomir2013b}\\ 
         Kepler-10 b &      0.837 &       4.60$\pm$1.26       &       1.46$\pm$0.02       &   3700 &                      \citet{Batalha2011} &                      \citet{Batalha2011}\\ 
         $^c$Kepler-11 b &     10.304 &       1.90$\pm$1.20       &       1.80$\pm$0.04       &    130 &                     \citet{Lissauer2011} &                     \citet{Lissauer2013}\\ 
         $^c$Kepler-11 c &     13.024 &       2.90$\pm$2.20       &       2.87$\pm$0.06       &     91 &                     \citet{Lissauer2011} &                     \citet{Lissauer2013}\\ 
         $^c$Kepler-11 d &     22.684 &       7.30$\pm$1.10       &       3.12$\pm$0.07       &     44 &                     \citet{Lissauer2011} &                     \citet{Lissauer2013}\\ 
         $^c$Kepler-11 f &     46.689 &       2.00$\pm$0.80       &       2.49$\pm$0.06       &     17 &                     \citet{Lissauer2011} &                     \citet{Lissauer2013}\\ 
         Kepler-18 b &      3.505 &       6.90$\pm$3.48       &       2.00$\pm$0.10       &    460 &                      \citet{Borucki2011} &                      \citet{Cochran2011}\\ 
         Kepler-20 b &      3.696 &       8.47$\pm$2.12       &       1.91$\pm$0.16       &    350 &                      \citet{Borucki2011} &                      \citet{Gautier2012}\\ 
         Kepler-20 c &     10.854 &      15.73$\pm$3.31       &       3.07$\pm$0.25       &     82 &                      \citet{Borucki2011} &                      \citet{Gautier2012}\\ 
         Kepler-20 d &     77.612 &       7.53$\pm$7.22       &       2.75$\pm$0.23       &      6.0 &                      \citet{Borucki2011} &                      \citet{Gautier2012}\\ 
         $^c$Kepler-30 b &	 29.334 &  	 11.3$\pm$1.4   &	 3.90	$\pm$0.20	&  21 &	 		\citet{Borucki2011} &		 \cite{Sanchis-Ojeda2012}\\
         $^c$Kepler-36 b &     13.840 &       4.46$\pm$0.30       &       1.48$\pm$0.03       &    220 &                      \citet{Borucki2011} &                       \citet{Carter2012}\\ 
         $^c$Kepler-36 c &     16.239 &       8.10$\pm$0.53       &       3.68$\pm$0.05       &    180 &                       \citet{Carter2012} &                       \citet{Carter2012}\\ 
         Kepler-68 b &      5.399 &       8.30$\pm$2.30       &       2.31$\pm$0.03       &    410 &                      \citet{Borucki2011} &                    \citet{Gilliland2013}\\ 
         Kepler-68 c &      9.605 &       4.38$\pm$2.80       &       0.95$\pm$0.04       &    190 &                      \citet{Batalha2013} &                    \citet{Gilliland2013}\\ 
           Kepler-78 b &      0.354 &       1.69$\pm$0.41       &       1.20$\pm$0.09       &   3100 &              \citet{Sanchis-Ojeda2013} &              \citet{Howard2013Nature}\\ 
           Kepler-100 c &     12.816 &       0.85$\pm$4.00       &       2.20$\pm$0.05       &    210 &                      \citet{Borucki2011} &                        \citet{Marcy2014}\\ 
           Kepler-100 b &      6.887 &       7.34$\pm$3.20       &       1.32$\pm$0.04       &    470 &                      \citet{Borucki2011} &                        \citet{Marcy2014}\\ 
           Kepler-100 d &     35.333 &      -4.36$\pm$4.10       &       1.61$\pm$0.05       &     56 &                      \citet{Borucki2011} &                        \citet{Marcy2014}\\ 
           Kepler-93 b &      4.727 &       2.59$\pm$2.00       &       1.50$\pm$0.03       &    220 &                      \citet{Borucki2011} &                        \citet{Marcy2014}\\ 
           Kepler-102 e &     16.146 &       8.93$\pm$2.00       &       2.22$\pm$0.07       &     17 &                      \citet{Borucki2011} &                        \citet{Marcy2014}\\ 
           Kepler-102 d &     10.312 &       3.80$\pm$1.80       &       1.18$\pm$0.04       &     31 &                      \citet{Borucki2011} &                        \citet{Marcy2014}\\ 
           Kepler-102 f &     27.454 &       0.62$\pm$3.30       &       0.88$\pm$0.03       &      8.3 &                      \citet{Borucki2011} &                        \citet{Marcy2014}\\ 
           Kepler-102 c &      7.071 &      -1.58$\pm$2.00       &       0.58$\pm$0.02       &     51 &                      \citet{Borucki2011} &                        \citet{Marcy2014}\\ 
           Kepler-102 b &      5.287 &       0.41$\pm$1.60       &       0.47$\pm$0.02       &     78 &                      \citet{Borucki2011} &                        \citet{Marcy2014}\\ 
          Kepler-94 b &      2.508 &      10.84$\pm$1.40       &       3.51$\pm$0.15       &    210 &                      \citet{Borucki2011} &                        \citet{Marcy2014}\\ 
          Kepler-103 b &     15.965 &      14.11$\pm$4.70       &       3.37$\pm$0.09       &    120 &                      \citet{Borucki2011} &                        \citet{Marcy2014}\\ 
          Kepler-106 c &     13.571 &      10.44$\pm$3.20       &       2.50$\pm$0.32       &     84 &                      \citet{Borucki2011} &                        \citet{Marcy2014}\\ 
          Kepler-106 e &     43.844 &      11.17$\pm$5.80       &       2.56$\pm$0.33       &     16 &                      \citet{Borucki2011} &                        \citet{Marcy2014}\\ 
          Kepler-106 b &      6.165 &       0.15$\pm$2.80       &       0.82$\pm$0.11       &    240 &                      \citet{Borucki2011} &                        \citet{Marcy2014}\\ 
         Kepler-106 d &     23.980 &      -6.39$\pm$7.00       &       0.95$\pm$0.13       &     43 &                      \citet{Borucki2011} &                        \citet{Marcy2014}\\ 
          Kepler-95 b &     11.523 &      13.00$\pm$2.90       &       3.42$\pm$0.09       &    180 &                      \citet{Borucki2011} &                        \citet{Marcy2014}\\ 
          Kepler-109 b &      6.482 &       1.30$\pm$5.40       &       2.37$\pm$0.07       &    440 &                      \citet{Borucki2011} &                        \citet{Marcy2014}\\ 
          Kepler-109 c &     21.223 &       2.22$\pm$7.80       &       2.52$\pm$0.07       &     95 &                      \citet{Borucki2011} &                        \citet{Marcy2014}\\ 
          Kepler-48 b &      4.778 &       3.94$\pm$2.10       &       1.88$\pm$0.10       &    170 &                      \citet{Borucki2011} &                        \citet{Marcy2014}\\ 
          Kepler-48 c &      9.674 &      14.61$\pm$2.30       &       2.71$\pm$0.14       &    230 &                      \citet{Borucki2011} &                        \citet{Marcy2014}\\ 
         Kepler-48 d &     42.896 &       7.93$\pm$4.60       &       2.04$\pm$0.11       &     14 &                      \citet{Borucki2011} &                        \citet{Marcy2014}\\ 
	Kepler-79 b	   & 	13.4845 &	 	10.9$\pm$6.70	  &	3.47$\pm$0.07 &	160 &	 	\citet{Borucki2011}	& 			\citet{Jontof-Hutter2013}\\
	Kepler-79 c	   & 	27.4029 &	 	5.9$\pm$2.10	 & 	3.72$\pm$0.08	 &	 63 &			 \citet{Borucki2011}	 &			 \citet{Jontof-Hutter2013}\\
	Kepler-79 e	   & 	81.0659 &	 	4.1$\pm$1.15	 & 	3.49$\pm$0.14	 &	 15 &	 		\citet{Borucki2011}	 &			 \citet{Jontof-Hutter2013}\\
          Kepler-113 c &      8.925 &      -4.60$\pm$6.20       &       2.19$\pm$0.06       &     51 &                      \citet{Borucki2011} &                        \citet{Marcy2014}\\ 
          Kepler-113 b &      4.754 &       7.10$\pm$3.30       &       1.82$\pm$0.05       &     64 &                      \citet{Borucki2011} &                        \citet{Marcy2014}\\ 
          Kepler-25 b &      6.239 &       9.60$\pm$4.20       &       2.71$\pm$0.05       &    670 &                      \citet{Borucki2011} &                        \citet{Marcy2014}\\ 
          Kepler-37 d &     39.792 &       1.87$\pm$9.08       &       1.94$\pm$0.06       &      7.7 &                      \citet{Borucki2011} &                        \citet{Marcy2014}\\ 
          Kepler-37 c &     21.302 &       3.35$\pm$4.00       &       0.75$\pm$0.03       &     16 &                      \citet{Borucki2011} &                        \citet{Marcy2014}\\ 
          Kepler-37 b &     13.367 &       2.78$\pm$3.70       &       0.32$\pm$0.02       &     37 &                      \citet{Borucki2011} &                        \citet{Marcy2014}\\ 
          Kepler-68 b &      5.399 &       5.97$\pm$1.70       &       2.33$\pm$0.02       &    380 &                      \citet{Borucki2011} &                        \citet{Marcy2014}\\ 
          Kepler-68 c &      9.605 &       2.18$\pm$3.50       &       1.00$\pm$0.02       &    220 &                      \citet{Borucki2011} &                        \citet{Marcy2014}\\ 
          Kepler-96 b &     16.238 &       8.46$\pm$3.40       &       2.67$\pm$0.22       &     74 &                      \citet{Borucki2011} &                        \citet{Marcy2014}\\ 
          Kepler-131 b &     16.092 &      16.13$\pm$3.50       &       2.41$\pm$0.20       &     72 &                      \citet{Borucki2011} &                        \citet{Marcy2014}\\ 
          Kepler-131 c &     25.517 &       8.25$\pm$5.90       &       0.84$\pm$0.07       &     29 &                      \citet{Borucki2011} &                        \citet{Marcy2014}\\ 
	Kepler-97 b &      2.587 &       3.51$\pm$1.90       &       1.48$\pm$0.13       &    850 &                      \citet{Borucki2011} &                        \citet{Marcy2014}\\ 
          Kepler-98 b &      1.542 &       3.55$\pm$1.60       &       1.99$\pm$0.22       &   1600 &                      \citet{Borucki2011} &                        \citet{Marcy2014}\\ 
          Kepler-99 b &      4.604 &       6.15$\pm$1.30       &       1.48$\pm$0.08       &     90&                      \citet{Borucki2011} &                        \citet{Marcy2014}\\ 
          $^d$Kepler-406 b &      2.426 &       4.71$\pm$1.70       &       1.43$\pm$0.03       &    710 &                      \citet{Borucki2011} &                        \citet{Marcy2014}\\ 
          $^d$Kepler-406 c &      4.623 &       1.53$\pm$2.30       &       0.85$\pm$0.03       &    290 &                      \citet{Borucki2011} &                        \citet{Marcy2014}\\ 
         Kepler-407 b &      0.669 &       0.06$\pm$1.20       &       1.07$\pm$0.02       &   3600 &                      \citet{Borucki2011} &                        \citet{Marcy2014}\\ 
         Kepler-409 b &     68.958 &       2.69$\pm$6.20       &       1.19$\pm$0.03       &      6.2 &                      \citet{Borucki2011} &                        \citet{Marcy2014}\\ 
         KOI-94 b &      3.743 &       10.50$\pm$4.60       &       1.71$\pm$0.16       &   1200 &                        \citet{Batalha2013} &                        \citet{Weiss2013}\\ 
         KOI-1612.01 &      2.465 &       0.48$\pm$3.20       &       0.82$\pm$0.03       &   1700 &                      \citet{Borucki2011} &                        \citet{Marcy2014}\\ 
         KOI-314 b &      0.669 &       0.06$\pm$1.20       &       1.07$\pm$0.02       &   3600 &                      \citet{Borucki2011} &                        \citet{Kipping2014}\\ 
         KOI-314 c &      0.669 &       0.06$\pm$1.20       &       1.07$\pm$0.02       &   3600 &                      \citet{Borucki2011} &                        \citet{Kipping2014}\\ 
\enddata

\tablenotetext{$^a$}{Incident stellar flux is calculated as $F/\fearth = (\rstar/\rsun)^2 (\teff/5778 \mathrm{K})^4 a^{-2} \sqrt{1/(1-e^2)}$, where $a$ is the semi-major axis in A.U. and $e$ is the eccentricity.  Typical errors are 10\%.}
\tablenotetext{$^b$}{Mass is from \citet{Endl2012}, radius is from \citet{Dragomir2013a}.  The density is calculated from these values.}
\tablenotetext{$^c$}{Planet mass determined by TTVs of a neighboring planet}
\tablenotetext{$^d$}{Planet mass and density updated based on additional RVs.}



\label{tab:mrf}

\end{deluxetable*}

\subsection{Inclusion of Mass Non-Detections}
For small exoplanets, uncertainties in the mass measurements can be of order the planet mass.  Although one might advocate for only studying planets with well-determined ($>3\sigma$) masses, imposing a significance criterion will bias the sample toward more massive planets at a given radius.  This bias is especially pernicious for small planets, for which the planet-induced RV signal can be small ($\sim1\ms$) compared to the noise from stellar activity ($\sim2\ms$) and Poisson photon noise ($\sim2\ms$).  We must include the marginal mass detections and non-detections in order to minimize bias in planet masses at a given radius.  

\citet{Marcy2014} employ a new technique for including non-detections.  They allow  a negative semi-amplitude in the Keplerian fit to the RVs and report the peak and 68$^{th}$ percentiles of the posterior distribution of the semi-amplitude.  The posterior distribution peak often corresponds to a ``negative" planet mass, although the wings of the posterior distribution encompass positive values.  Although planets cannot have negative masses in nature, random fluctuations in the RVs from noise can produce a velocity curve that is low when it should be high, and high when it should be low, mimicking the RV signature of a planet 180$^\circ$ out of phase with the transit-determined ephemeris.  Since the planetary ephemeris is fixed by the transit, \citet{Marcy2014} allow these cases to be fit with a negative semi-amplitude solution in their MCMC analysis.  Reporting the peak of the posterior distribution is statistically meaningful because there are also cases where the fluctuations in RVs from the random noise happen to correlate with the planetary signal, artificially increasing the planet mass.  We include non-detections (as negative planet masses and low-significance positive planet masses) to avoid statistical bias toward large planet masses at a given radius.

Including literature values, which typically only report planet mass if the planet mass is detected with high confidence, slightly biases our sample toward higher masses at a given radius.  We include the literature values to provide a larger sample of exoplanets.

\section{The Mass-Radius Relations}
In Figure \ref{fig:rm_4}, we show the measured planet densities and planet masses for $\rpl < 4 \rearth$.  In addition, we show the weighted mean planet density and mass in bins of 0.5 \rearth.  The weighted mean densities and masses guide the eye, demonstrating how the ensemble density and mass change with radius.  We also include the solar system planets.  Examining the solar system terrestrial planets and the weighted mean density at 1.5 \rearth, we see that planet density increases with increasing radius up to 1.5 \rearth.  For planets between 1.0 and 1.5 \rearth, the weighted mean density achieves a maximum at $7.6\pm1.2 \gcc$, and the weighted center of the bin is at 1.4 \rearth.  Above 1.5 \rearth, planet density decreases with increasing radius.  The break in the density-radius relation motivates us to explore different empirical relations for planets smaller and larger than 1.5 \rearth.

Exoplanets smaller than 1.5 \rearth\ mostly have mass uncertainties of order the planet mass, except for Kepler-10 b, Kepler-36 b, Kepler-78 b, and Kepler-406 b (KOI-321 b).  Because there are so few planets with well-determined masses in this regime, we include the terrestrial solar system planets (Mercury, Venus, Earth, Mars) in a fit to the planets smaller than 1.5 \rearth.  We impose uncertainties of 20\% in their masses and 10\% in their radii so that the solar system planets will contribute to, but not dominate, the fit.  Because the solar system planets appear to satisfy a linear relation between density and radius, we choose a linear fit to planet density vs. radius.  We find:
\begin{equation}
\rhopl = 2.43 + 3.39 \left(\frac{\rpl}{\rearth}\right) \gcc.
\label{eqn:dr_lin_rocky}
\end{equation}
Transforming the predicted densities to masses via 
\begin{equation}
\frac{\mpl}{\mearth} = \left(\frac{\rhopl}{\rho_\oplus}\right) \left(\frac{\rpl}{\rearth}\right)^3
\end{equation}
and calculating the residuals with respect to the measured planet masses, we obtain reduced $\chi^2 = 1.3$, RMS$=$2.7 \mearth.

For exoplanets satisfying $1.5 \le \rpl/\rearth < 4$, we calculate an empirical fit to their masses and radii, yielding:
\begin{equation}
\frac{\mpl}{\mearth} = 2.69 \left(\frac{\rpl}{\rearth}\right)^{0.93}
\label{eqn:mr_plaw}
\end{equation}
with reduced $\chi^2=3.5$ and RMS=4.7 \mearth.  We exclude Uranus and Neptune from this fit because they differ from the exoplanets in our sample.  Most of the exoplanets in our sample have $P < 50$ days, and so we do not expect them to resemble Uranus and Neptune, which have orbital periods of tens of thousands of days.

The empirical density- and mass-radius relations and their goodness of fit are summarized in Table \ref{tab:mr_relations}.  Below, we discuss the implications of these relations for planet compositions.


\section{Implications for Planet Compositions}
	
\subsection{Interpretation of the Density-Radius and Mass-Radius Relations}
The peak of the density-radius relation at 1.5 \rearth\ and 8 \gcc\ is consistent with the \citet{Seager2007} prediction for the density of a 1.5 \rearth\ Earth-composition planet.  The density peak at 1.5 \rearth\ is also consistent with the division between rocky and non-rocky planets determined in \citet{Rogers2014} through Bayesian modeling and MCMC analysis.  Following the \citet{Seager2007} prediction for the density of an Earth-composition (67.5\% MgSiO$_3$, 32.5\% Fe) planet, we see a predicted increase in planet density with increasing planet radius.  This is because rock is slightly compressible, causing an increase in density with increasing planet radius.   Because the compression of rock is slight, we could, in principle, take a first-order Taylor expansion to the equation of state of a rocky planet and approximate that density increases linearly with radius; this is consistent with our empirical, linear density-radius fit.  If the exoplanets in this regime are indeed rocky, our inclusion of the solar system planets is justified because the orbital period (out to Earth's orbit) and incident flux on a rocky planet should have very little effect on that planet's mass and radius.  Equation \ref{eqn:dr_lin_rocky} and the density-radius relation from \citet{Seager2007} are both consistent with the interpretation that planets smaller that 1.5 \rearth\ are rocky, but Equation \ref{eqn:dr_lin_rocky} has advantages in that it (a) is empirical, and (b) passes closer to Earth, Venus, and Mars, which are known to be rich in silicon and magnesium (unlike Mercury, which is iron-rich).  Additional and more precise, mass measurements for planets smaller than 1.5 \rearth\ are necessary to hone the density-radius relation below 1.5\rearth\ and examine any scatter about the relation.

For planets between 1.5 and 4 \rearth, the weighted mean density decreases with increasing planet radius, making these planets inconsistent with a rocky composition.  The decrease in density must be due to an increasing fraction of volatiles, which we argue must be at least partially in the form of H/He envelopes.  The gentle rise in planet mass with increasing radius indicates a substantial change in volume (from 3.4 to 64 times the volume of Earth) for very little change in mass (from 4 to 10 Earth masses; see Figure \ref{fig:rm_4}).  A water layer alone cannot explain this enormous change in volume for so little added mass; lightweight gas must be present in increasing quantities with increasing planetary radius.  We can do a thought experiment: if exoplanets at 4 \rearth\ (which have densities of about 1 \gcc) are made entirely of water, what distinguishes them from the 1.5 \rearth\ planets whose compositions are likely rocky?  We see no empirical evidence in the weighted mean density vs. radius that would suggest a shift between planets that do have rocky cores at 1.5 \rearth, and planets that do not have rocky cores at 4 \rearth.  Moreover, planet formation theory makes it very likely that water content will be accompanied by at least a comparable amount of silicates and iron-nickel.  The smooth decline in the weighted mean density from 1.5 to 4 \rearth\ seems more consistent with the accumulation of lightweight gaseous envelopes upon rocky cores.

\subsection{Scatter about the Relations}
The moderate reduced $\chi^2$ (6.3) to the mass-radius relation between 1.5 and 4 \rearth\ indicates that measurement errors do not explain the variation in planet mass at a given radius.  Only a diversity of planet compositions at given radius explains the large scatter in planet mass.  Perhaps the mass diversity at a given radius results from different core masses in planets with large gaseous envelopes \citep[as argued in ][]{Lopez2013}; the size of the planet is determined by the fraction of gas, but the mass is determined by the size of the rocky core.   In addition, water layers between the rocky cores and gaseous envelopes could help account for the scatter in mass at a given radius.

\subsection{Previous Studies of the Mass-Radius Relation}
\citet{Lissauer2011}, \citet{Enoch2012}, \citet{Kane2012}, and \citet{Weiss2013} suggest that the mass-radius relation is more like $\mpl \propto \rpl^2$ for small exoplanets.  However, these studies include Saturn or Saturn-like planets at the high-mass end of their ``small planet" populations.  Such planets are better described as part of the giant planet population and are not useful in determining an empirical mass-radius relation of predictive power for small exoplanets.

In a study of planets with $\mpl < 20\mearth$, \citet{WL2013} find $\mpl/\mearth = 3 \rpl/\rearth$ in a sample of 22 pairs of planets that exhibit strong anti-correlated TTVs in the \textit{Kepler} data.  Our independent assessment of 65 exoplanets, 52 of which are not analyzed in \citet{WL2013}, is consistent with this result for planets larger than 1.5\rearth.  \citet{WL2013} note that a linear relation between planet mass and radius is dimensionally consistent with a constant escape velocity from the planet (i.e. $v_{\mathrm{esc}}^2 \sim \mpl/\rpl$).  The linear mass-radius relation might result from photo-evaporation of the atmospheres of small planets near their stars \citep{Lopez2012}.

\subsection{Masses from TTVs are Lower than Masses from RVs}
\begin{figure*}[htbp] 
   \centering
    \includegraphics[width=7in]{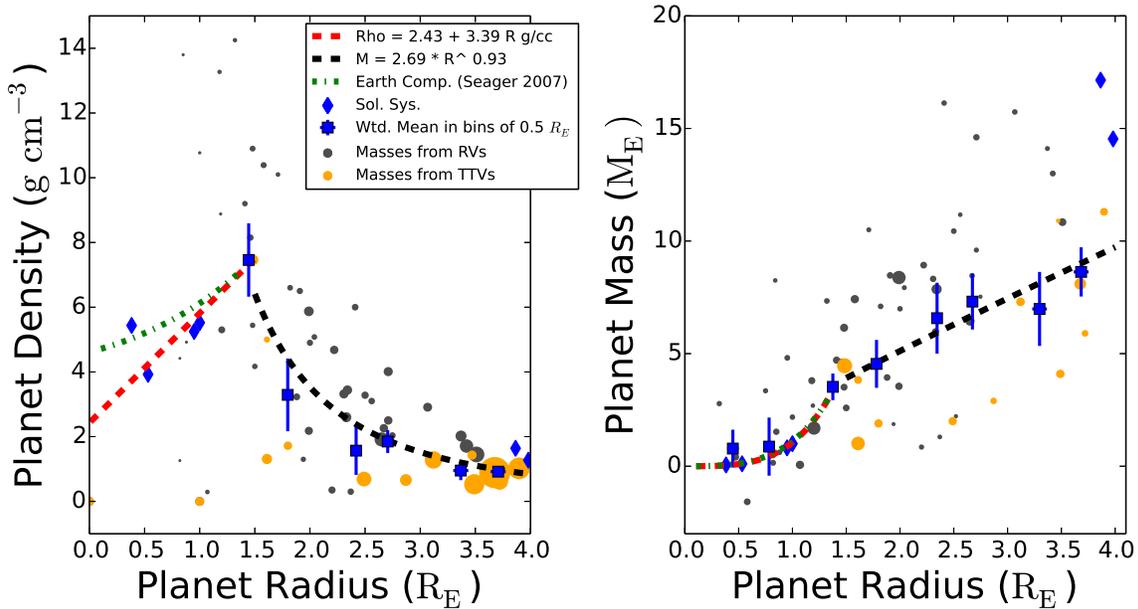} 
   \caption{\small \textbf{Left:} Density vs. radius for 65 exoplanets.  Gray points  have RV-determined masses; orange points have TTV-determined masses, and the point size corresponds to $1/\sigma(\rhopl)$.  The blue squares are weighted mean densities in bins of 0.5 \rearth, with error bars representing $\sigma_i/\sqrt{N_i}$, where $\sigma_i$ is the standard deviation of the densities and $N_i$ is the number of exoplanets in bin $i$.  We omit the weighted mean densities below 1.0 \rearth\ because the scatter in planet densities is so large that the error bars span the range of physical densities (0 to 10 \gcc).  The blue diamonds indicate solar system planets.  The red line is an empirical density-radius fit for planets smaller than 1.5\rearth, including the terrestrial solar system planets.  The green line is the mass-radius relation from \citet{Seager2007} for planets of Earth composition (67.5\% MgSiO$_3$, 32.5\% Fe).  The increase in planet density with radius for $\rpl < 1.5 \rearth$ is consistent with a population of rocky planets.  Above 1.5 \rearth, planet density decreases with planet radius, indicating that as planet radius increases, so does the fraction of gas.  \textbf{Right:} Mass vs. radius for 65 exoplanets.  Same as left, but the point size corresponds to $1/\sigma(\mpl)$, and the blue squares are the weighted mean masses in bins of $0.5 \rearth$, with error bars representing $\sigma_i/\sqrt{N_i}$, where $\sigma_i$ is the standard deviation of the masses and $N_i$ is the number of exoplanets in bin $i$. The black line is an empirical fit to the masses and radii above 1.5 \rearth; see equation \ref{eqn:mr_plaw}.  The weighted mean masses were not used in calculating the fit.   Some mass and density outliers are excluded from these plots, but are included in the fits. }
   \label{fig:rm_4}
\end{figure*}

We have included planets with masses determined by TTVs in Table \ref{tab:mrf}, Figure \ref{fig:rm_4} and the mass-radius relations.  The TTV masses included in this work are the result of dynamical modeling that reproduces the observed TTV signatures in the Kepler light curve.  Planets with TTV-determined masses are marked with superscript $c$ in Table \ref{tab:mrf}.  In Figure \ref{fig:rm_4}, the TTV planets are shown as orange points; they are systematically less massive than the RV-discovered planets of the same radii \citep[also see][]{Jontof-Hutter2013}.  A T-test comparing the residual masses from the RVs to the TTVs results in a two-tailed P-value of 0.03, indicating the two samples, if drawn from the same distribution, would be this discrepant 3\% of the time.  An empirical fit between only the RV-determined planet masses and their radii for $1.5 < \rpl/\rearth < 4$ yields a similar solution to equation \ref{eqn:mr_plaw}, but predicts slightly higher masses and less scatter of the residuals: $\mpl/\mearth = 4.87 \left( \rpl/\rearth \right)^{0.63}$, reduced $\chi^2 = 2.0.$

The systematic difference between the TTV and RV masses is unlikely to stem from a bias in the RVs.  Either the TTVs are systematically underestimating planet masses (possibly because other planets in the system damp the TTVs), or compact systems amenable to detection through TTVs have lower-density planets than non-compact systems \citep[e.g. the Kepler-11 system,][]{Lissauer2013}.  That \citet{WL2013} also find $\mpl/\mearth \approx 3 \left(\rpl/\rearth\right)$ suggests that the TTV masses might be reliably systematically lower, although \citet{WL2013} use analytic rather than numerical methods to estimate planet masses.

\subsection{Absence of Strong Correlations to Residuals}
We investigate how the residual mass correlates with various orbital properties and physical properties of the star.  The residual mass is the measured minus predicted planet mass at a given radius.  The quantities we correlate against are: planet orbital period, planet semi-major axis, the incident flux from the star on the planet, stellar mass, stellar radius, stellar surface gravity, stellar metallicity, stellar age, and stellar velocity times the sine of the stellar spin axis inclination (which are obtained through exoplanets.org or the papers cited in Table \ref{tab:mrf}).  In these data, the residual mass does not strongly correlate with any of these properties.

We find possible evidence of a correlation between residual planet mass and stellar metallicity for planets smaller than 4\rearth.  The Pearson R-value of the correlation is 0.25, resulting in a probability of 7\% that the residual planet mass and stellar metallicity are not correlated, given the residual masses and metallicities.  However, given that we looked for correlations among 9 pairs of variables, the probability of finding a $93.6\%$ confidence correlation in any of the 9 trials due to random fluctuation is $1 - 0.936^9 = 0.45$, meaning there is only a 55\% chance that the apparent metallicity correlation is real.

\begin{deluxetable*}{llll}
\tablewidth{0pt} 
\tablecaption{Empirical Mass-Radius and Density-Radius Relations}
\tablenum{2}
\tablehead{\colhead{Planet Size} & \colhead{Equation} & \colhead{Reduced $\chi^2$} & \colhead{RMS}} 

\startdata
$^{a}\rpl < 1.5 \rearth$ &  ${\rhopl} = 2.43 + 3.39 \left(\frac{\rpl}{\rearth}\right) \gcc$ & 1.3 & 2.7 \mearth \\
$1.5 \le \rpl/\rearth< 4$ &  $\frac{\mpl}{\mearth} = 2.69\left(\frac{\rpl}{\rearth}\right)^{0.93}$ & 6.2 & 4.3 \mearth \\
\enddata
\tablenotetext{a}{Including terrestrial solar system planets Mercury, Venus, Earth, and Mars.}


\label{tab:mr_relations}

\end{deluxetable*}


\section{Conclusions}
The weighted mean exoplanet density peaks at approximately 1.4 \rearth\ and 7.6 \gcc\, which is consistent with an Earth-composition planet.  Planet density increases with radius up to 1.5 \rearth, but above 1.5 \rearth, planet density decreases with planet radius. Planets smaller than 1.5\rearth\ are consistent with a linear density-radius relation, and are also consistent with the \citet{Seager2007} Earth composition curve.  Above 1.5\rearth, the decrease in planet density with increasing radius can only be due to the inclusion of volatiles, and so planets larger than 1.5 \rearth\ are generally inconsistent with a purely rocky composition.  Among planets larger than 1.5\rearth, the gentle rise in planet mass with increasing radius indicates a substantial change in radius for very little change in mass, suggesting that lightweight H/He gas is present in increasing quantities with increasing planetary radius.


\begin{figure*}[htbp] 
   \centering
    \includegraphics[width=7in]{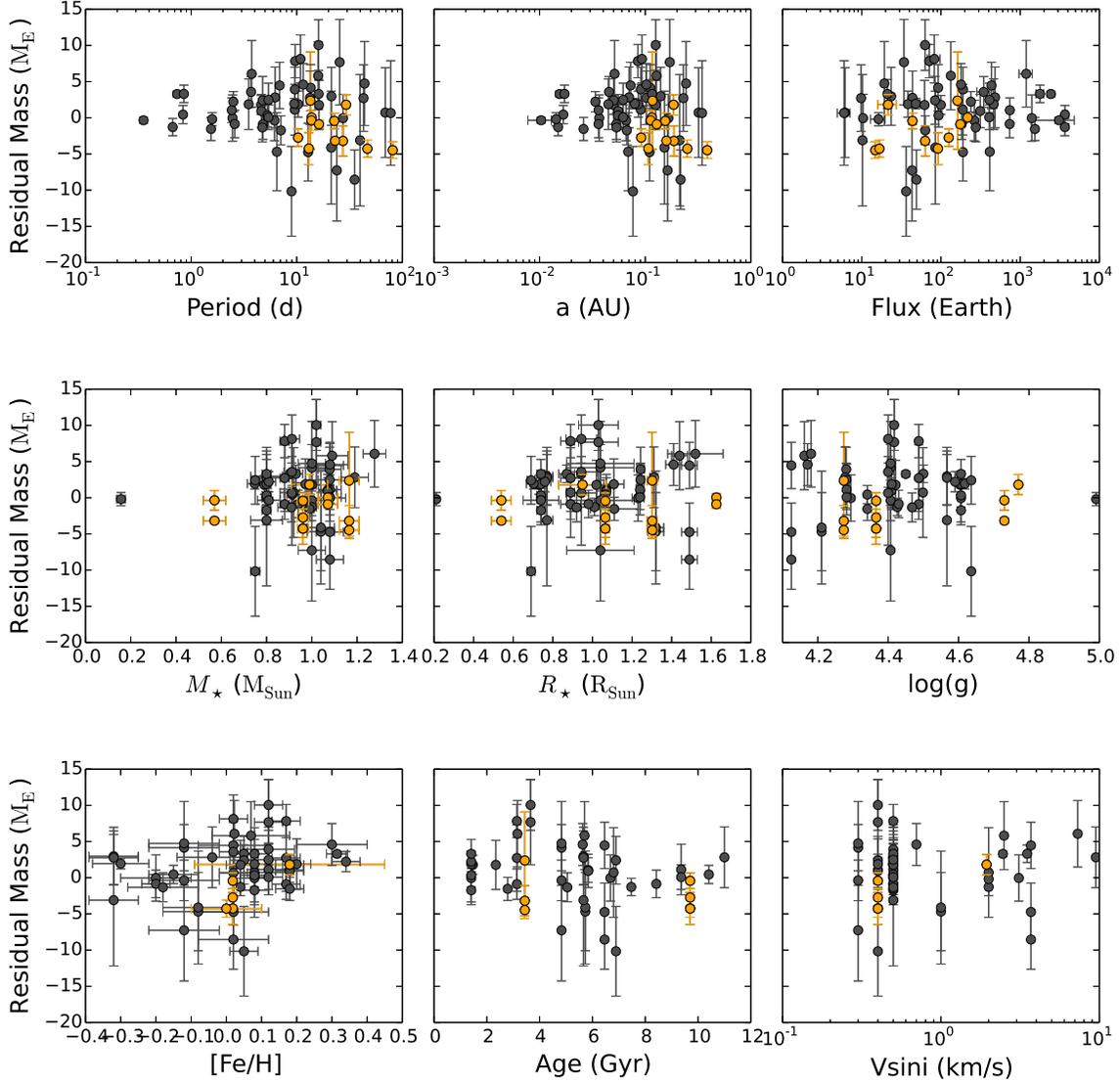} 
   \caption{\small Mass residuals (measured minus the mass predicted from equations \ref{eqn:dr_lin_rocky} - \ref{eqn:mr_plaw}) versus (top left to bottom right): planet orbital period, planet semi-major axis, incident stellar flux, stellar mass, stellar radius, surface gravity, metallicity (compared to solar), stellar age, and stellar $v\mathrm{sin}i$. Error bars are 1$\sigma$ uncertainties, and the orange points are residuals of the TTV-determined masses.  None of the residuals show a significant correlation, although more mass measurements might elucidate a correlation with metallicity.}
   \label{fig:resids}
\end{figure*}




\end{document}